\documentclass[preprint]{aastex}

\shorttitle{Detection of Circular Polarization in M81}
\shortauthors{Brunthaler et al.}

\begin{document}

\title{Detection of Circular Polarization in M81*}

\author{A. Brunthaler\altaffilmark{1,2}, G.C. Bower\altaffilmark{3},
        H. Falcke\altaffilmark{2}, \& R. R. Mellon\altaffilmark{4}}

\altaffiltext{1}{Harvard-Smithsonian Center for Astrophysics, 60 Garden Street MS 42, Cambridge, MA 02138; abrunthaler@cfa.harvard.edu}
\altaffiltext{2}{Max-Planck-Institut f\"ur Radioastronomie, Auf dem H\"ugel 69, 53121 Bonn, Germany; hfalcke@mpifr-bonn.mpg.de}
\altaffiltext{3}{Radio Astronomy Lab, UC Berkeley, 601 Campbell Hall,  Berkeley CA 94720}
\altaffiltext{4}{Department of Astronomy and Astrophysics, 525 Davey Lab, The Pennsylvania State University, University Park, PA 16802}

\begin{abstract}
We report the detection of circular polarization in the compact radio
jet of the nearby spiral galaxy M81 (M81*). The observations were made
with the Very Large Array at 4.8 and 8.4 GHz and circular polarization
was detected at both frequencies. We estimate a value of
$m_{c}=0.54\pm0.06\pm0.07\%$ at 8.4 GHz and $m_{c}=0.27\pm0.06\pm0.07\%$ 
at 4.8 GHz for the
fractional circular polarization. The errors are separated into statistical 
and systematic terms. The spectrum of the circular
polarization is possibly inverted which would be unusual for AGN.  
We also detected
no linear polarization in M81* at a level of 0.1\% implying that the
source has a very high circular-to-linear polarization ratio as found so
far only in Sgr~A*, the central radio source in our Galaxy. This
further supports the idea that M81* is a scaled-up version of Sgr A*
and suggests that the polarization properties are intrinsic to the two
sources and are not caused by a foreground screen in the Galaxy.
\end{abstract}

\keywords{galaxies: active, galaxies: individual (M81), polarization}

\section{Introduction}

The nearby spiral galaxy M81 (NGC~3031) is very similar to our own
Galaxy in many ways. It resembles the Milky Way in type, size and mass.
It also contains a radio core, M81*, that is most likely associated with
a supermassive black hole.

Previous VLBI (Bietenholz et al. 1996), multiwavelength (Ho, Filippenko, and Sargent
1996) and submm observations (Reuter \& Lesch
1996) have shown that M81* is very
similar to Sgr~A*, the central radio source in our Galaxy. A
comparison of the two radio sources therefore may provide interesting
insights in their nature.

Especially for Sgr A* the nature of the radio emission has been
debated for quite some time (see Melia \& Falcke 2001, for a
review). Either an origin in an accretion flow (Melia 1992, Narayan \&
Mahadevan 1995) or in a jet has been proposed (Falcke, Mannheim, and Biermann 1993, 
Falcke \& Markoff 2000). The jet model, within the context of
the jet-disk symbiosis, has also been applied to M81* (Falcke 1996)
where it can reproduce the radio flux density and the size of the 
radio core simply by
changing the accretion rate. Indeed, Bietenholz, Bartel, and Rupen (2000) have 
discovered a one-sided, though very compact jet in M81*
and they pointed out that in terms of power, jet length, and perhaps
accretion rate, M81* is intermediate between radio cores of quasars
and Sgr~A*.

Recently, Bower, Falcke, and Backer (1999a) detected circular
polarization in Sgr~A* in absence of linear polarization (Bower et
al. 1999b\&c). This was confirmed by Sault \&
Macquart (1999). This result is surprising, since linear
polarization usually exceeds circular by a large factor in AGN radio jets 
(Wardle et al. 1998; Rayner, Norris, and Sault 2000).
Bower et al. (1999d) proposed that low-energy electrons intrinsic to the
source reduce the linear polarization through Faraday de-polarization
and convert linear polarization into circular polarization (Pacholczyk
1977; Jones \& O'Dell 1977). This would provide very important information on the
matter content of Sgr A*. On the other hand, it cannot be
excluded that the accretion region around Sgr A* contributes to the unusual
polarization (Bower et al. 1999b, Quataert \& Gruzinov
2000).  Less likely, but not fully excluded is the possibility that
the hyperstrong scattering screen could induce the circular polarization
through propagation effects (Macquart \& Melrose 2000).

Since M81* is not in the direction of the Galactic Center and is 
likely to be observed under a very different inclination angle, 
observations of this closely related source are independent of the details
of the geometry of Sgr A*.  Here we report the results of circular
and linear polarization observations of M81* which indicate that indeed
M81* and Sgr A* have similar polarization properties.

\section{Observations and Results}

We observed M81* with the VLA on 2000 April 4 at 8.4 GHz in C
configuration; on 2000 September 27 at 4.8 GHz and at 8.4 GHz in D
configuration; and on 2000 November 4 at 4.8 GHz in A configuration
with a bandwidth of 50 MHz in right circular polarization (RCP) and
left circular polarization (LCP). We used 3C~48 as the primary flux
density calibrator, J1044+719 as secondary calibrator and J1053+704 as
check source. We went through the cycle J1044+719 - M81* - J1053+704 -
J1044+719 - M81* - J1053+704 - J1044+719 once for each frequency on
2000 September 27 and twice at 4.8 GHz on 2000 November 4. The
observation on 2000 April 4 was part of a polarization survey of 11
low luminosity AGNs of which M81 is the closest (Mellon et al. 2000).
Here we went through
the cycle J1044+719 - J1053+704 - M81* - J1044+719 five times. A
detailed discussion of this survey will be given in an upcoming paper
(Bower, Mellon, and Falcke 2001a).

Data reduction was performed with AIPS. Absolute flux densities were 
calibrated with
the source 3C~48. Then amplitude and phase self-calibration was performed on
J1044+719. This forces J1044+719 to have zero circular polarization. The 
amplitude calibration solutions were transfered to M81* and J1053+704. 
Finally each source was phase self-calibrated and imaged in Stokes
I and V. Flux densities were determined by fitting a beam-sized 
Gaussian at the image 
center. 

The Stokes parameter V is measured as the difference between the left- and
right-handed parallel polarization correlated visibilities, LL and RR. Errors
in circular polarization measurements with the VLA have numerous origins: 
thermal noise, gain errors, beam squint, second-order leakage corrections,
unknown calibrator polarization, background noise and radio frequency 
interference. The thermal noise is given by the rms in each map and the
values are shown in Table 1. The errors caused by amplitude calibration 
errors, beam squint and polarization leakage scale with the source strength 
and therefore the fractional circular polarization is a more relevant
indicator for the detection of circular polarization.
A detailed discussion of these errors is given in Bower et al. (1999a) and 
Bower et al. (2001b). We calculated the systematic errors based on the model 
for VLA circular polarization from Bower et al. 2001b. These values are given
in Table 1 and 2. 


For the observation on 2000 April 4, 3C~48 was used to calibrate the position
angle of linear polarization and the sources were also imaged in Stokes U and Q
to provide linear polarization information.

The results for M81*, the calibrator J1044+719 and the check-source J1053+704 
are given for 4.8 GHz in Table~\ref{tbl-1} and for 8.4 GHz in 
Table~\ref{tbl-2}. The check source J1053+704 showed weak circular polarization
at a level of $0.14\%$ on 2000 April 04 at 8.4 GHz. At all other observations,
the calibrator source and the check source showed no detectable circular 
polarization. The upper limits in Table 1 
and 2 were determined by fitting a beam-sized Gaussian at the image center. 
M81* showed circular polarization in all observations. 

We find a value of $m_{c}=0.27\pm0.06\pm0.07\%$ and 
$m_{c}=0.26\pm0.04\pm0.06\%$ for the
fractional circular polarization at 4.8 GHz on 2000 September 27 and
2000 November 4 respectively. At 8.4 GHz the degree of circular
polarization was $m_{c}=0.25\pm0.02\pm0.05\%$ on 2000 April 4 and 
$m_{c}=0.54\pm0.06\pm0.07\%$ on 2000 September 27. The errors are separated 
into statistical and systematic terms.
Fig.~\ref{fig1} shows the Stokes V map (contours)
and the total intensity (grey scale) of M81* on 2000 September 27 at
8.4 GHz. M81* showed no detectable linear polarization, with an upper
limit of $0.1\%$.


At 4.8 GHz, the measured values for circular polarization in M81* are 5 and 6
times the rms in the image for the observations on 2000 September 27 and
2000 November 4, while the upper limits on circular polarization for the check 
source are only 3 and 2 times the image rms for the two observations. The
detection is stronger in the later observation, since the observing time was
more than twice as long as in the first observation.

In our observation on 2000 September 27, the measured value for circular 
polarization in M81* at 8.4 GHz is 10 times the rms in the image for the 
observation on 2000 September 27, while the upper limit on circular 
polarization for the check source is 4 times the rms. On 2000 April 04, we 
detected circular polarization in both M81* and the check source at a level 
of 10 times the image rms, but the fractional circular polarization is a 
factor of 2 higher in M81*.

If the detections would have
been caused by calibration errors, beam squint, polarization leakage or unknown
calibrator polarization, fractional circular polarization should have been 
detected in the check source with the same value.

\section{Discussion}

The mechanism for the production of circular polarization in AGN is
still not known with absolute certainty and several mechanisms have
been proposed. An important new parameter, that has been largely
ignored so far, is the circular-to-linear polarization ratio $R_{\rm
CL}=m_{\rm c}/m_{\rm l}$. For powerful blazars one typically has
$R_{\rm CL}\ll1$ (Homan \& Wardle 1999) for the cores of jets. In the
radio cores of very low-power AGN, M81* and Sgr A*, we now find
$R_{\rm CL}\gg1$ with limits on the linear polarization that are very
low.

This first raises the question of how to reduce the linear
polarization to such undetectable levels. Already Bower et
al. (1999c\&d) suggested that Faraday de-polarization in the
interstellar medium, and specifically the scattering medium in the
Galactic Center, could not be responsible for this. Our finding of a
high $R_{\rm CL}=m_{\rm c}/m_{\rm l}$ in M81*, which is not scatter
broadened, confirms this notion. This also suggests that the
birefringent screen proposed by Macquart \& Melrose (2000) is not
applicable here. On the other hand, de-polarization by a hot,
geometrically thick accretion flow (Agol 2000, Quataert \& Gruzinov
2000) is not excluded but also does not yet explain the presence of
circular polarization.

A viable mechanism could be Faraday conversion (Pacholczyk 1977; Jones
\& O'Dell 1977) of linear polarization to circular polarization caused by 
the lowest energy relativistic electrons. Bower et al.
(1999a) proposed a simple model for Sgr~A* in which low-energy
electrons reduce linear polarization through Faraday de-polarization
and convert linear polarization into circular polarization. Conversion
also affects the spectral properties of circular polarization and may
lead to a variety of spectral indices, including inverted spectra
(Jones \& O'Dell 1977). In inhomogeneous sources, conversion can
produce relatively high fractional circular polarization (Jones 1988).

Of course, synchrotron radiation has a small intrinsic component of
circular polarization (Legg \& Westfold 1968) which can play an
important role. However, this intrinsic circular polarization will be
reduced by field reversals and optical depth effects. Finally,
gyro-synchrotron emission, can also lead to high circular polarization
with an inverted spectrum and low linear polarization (Ramaty 1969).
All the latter mechanisms are to some degree related and require that
M81* and Sgr A* both contain a rather large number of low-energy
electrons.

A more detailed discussion of this issue will be presented in an upcoming 
paper by Beckert et al. (2001).

\section{Summary and Conclusion}

We have presented VLA observations of M81* at 4.8 and 8.4 GHz, and
circular polarization was clearly detected at both frequencies with a
possible flat-to-inverted spectrum at a level of 0.25-0.5\%.

In most AGN the fractional circular polarization typically is
$m_{c}<0.1\%$ with only a few cases where $m_{c}$ approaches $0.5\%$
(Weiler \& de Pater 1983). The degree of circular polarization usually
peaks near 1.4 GHz and decreases strongly with increasing
frequency. Surprisingly, the fractional circular polarization in M81*
is higher at 8.4 GHz than at 4.8 GHz for our observation on 2000 September 27.
The value at 4.8 GHz showed no variation between two epoch separated by six 
weeks despite a change in total intensity.


We also found that M81* shows less than $0.1\%$ linear polarization
thus making M81* the second low-luminosity AGN radio core, after
Sgr~A*, with a circular-to-linear polarization $R_{\rm CL}\gg1$. This
confirms that both sources may be of similar nature and similar models
may apply. It also strongly suggests that the polarization properties
are indeed intrinsic to the two sources and are not caused by the
Galactic scattering screen. In this respect it is important that M81*
was resolved in to a jet-like structure by VLBI, indicating that jets
indeed can produce an $R_{\rm CL}$ as high as observed here and in
Sgr~A*. 

Since the basic mechanisms to explain circular polarization involve
low-energy electrons one will need to consider their origin and their
effects on the spectrum in future models. This will provide one with
an important new constraint for the sub-Eddington accretion onto
supermassive black holes. Even if the radio emission is produced in a
jet, the composition and temperature of this plasma should reflect the
plasma properties of the accretion flow very close to the black hole.

\begin{acknowledgements}
The National Radio
Astronomy Observatory is a facility of the National Science Foundation
operated under cooperative agreement by Associated Universities, Inc.
\end{acknowledgements}

\clearpage

\begin{table}
\begin{center}
\caption{Total flux density I, circularly polarized flux density $P_{c}$, image rms and fractional circular polarization $m_{c}$ at 4.8 GHz for the calibrator J1044+719, M81 and the check source J1053+704 with the statistical and calculated systematic errors.\label{tbl-1}}
\begin{tabular}{cccrrrrrrrr}
\tableline\tableline
Source &Date & Time on source&I [mJy] & $P_{c}$ [mJy]& rms [mJy]&$m_{c}~[\%]$&&\\
\tableline
J1044+719 & 2000 Sep. 27&3m20s&1443&$<0.92$&0.34&$<0.06$&$\pm0.02$&-\\
    & 2000 Nov. 04&5m17s&1503&$<0.51$&0.15&$<0.03$&$\pm0.01$&-\\
M81* & 2000 Sep. 27&5m03s& 193.5&0.53&0.11&0.27&$\pm0.06$&$\pm0.07$\\
    & 2000 Nov. 04&12m08s& 160.7&0.41&0.07&0.26&$\pm0.04$&$\pm0.06$\\
J1053+704 & 2000 Sep. 27&1m49s& 355.1&$<0.53$&0.17&$<0.15$&$\pm0.05$&$\pm0.07$\\
    & 2000 Nov. 04&4m06s& 362.1&$<0.26$&0.11&$<0.07$&$\pm0.03$&$\pm0.06$\\
\tableline
\end{tabular}
\end{center}
\end{table}

\begin{table}
\begin{center}
\caption{Total flux density I, circularly polarized flux density $P_{c}$, image rms and fractional circular polarization $m_{c}$ at 8.4 GHz for the calibrator J1044+719, M81 and the check source J1053+704 with the statistical and calculated systematic errors.\label{tbl-2}}
\begin{tabular}{cccrrrrrr}
\tableline\tableline
Source & Date &Time on source& I [mJy] & $P_{c}$ [mJy]& rms [mJy]&$m_{c}~[\%]$&&\\
\tableline
J1044+719 & 2000 Apr. 04 &12m08s& 1542&$<0.02$&0.03&$<0.001$&$\pm0.002$&-\\
& 2000 Sep. 27&3m28s&1316&$<0.60$&0.28&$<0.05$&$\pm0.02$&-\\
M81* & 2000 Apr. 04&4m59s& 263.6&0.66&0.06&0.25&$\pm0.02$&$\pm0.05$\\
& 2000 Sep. 27&5m03s& 193.6&1.05&0.11&0.54&$\pm0.06$&$\pm0.07$\\
J1053+704 & 2000 Apr. 04&10m01s&467.7&0.66&0.07&0.14&$\pm0.01$&$\pm0.05$\\
& 2000 Sep. 27&2m04s& 405.0&$<0.68$&0.19&$<0.17$&$\pm0.05$&$\pm0.07$\\
\tableline
\end{tabular}
\end{center}
\end{table}

\begin{figure}
\plotone{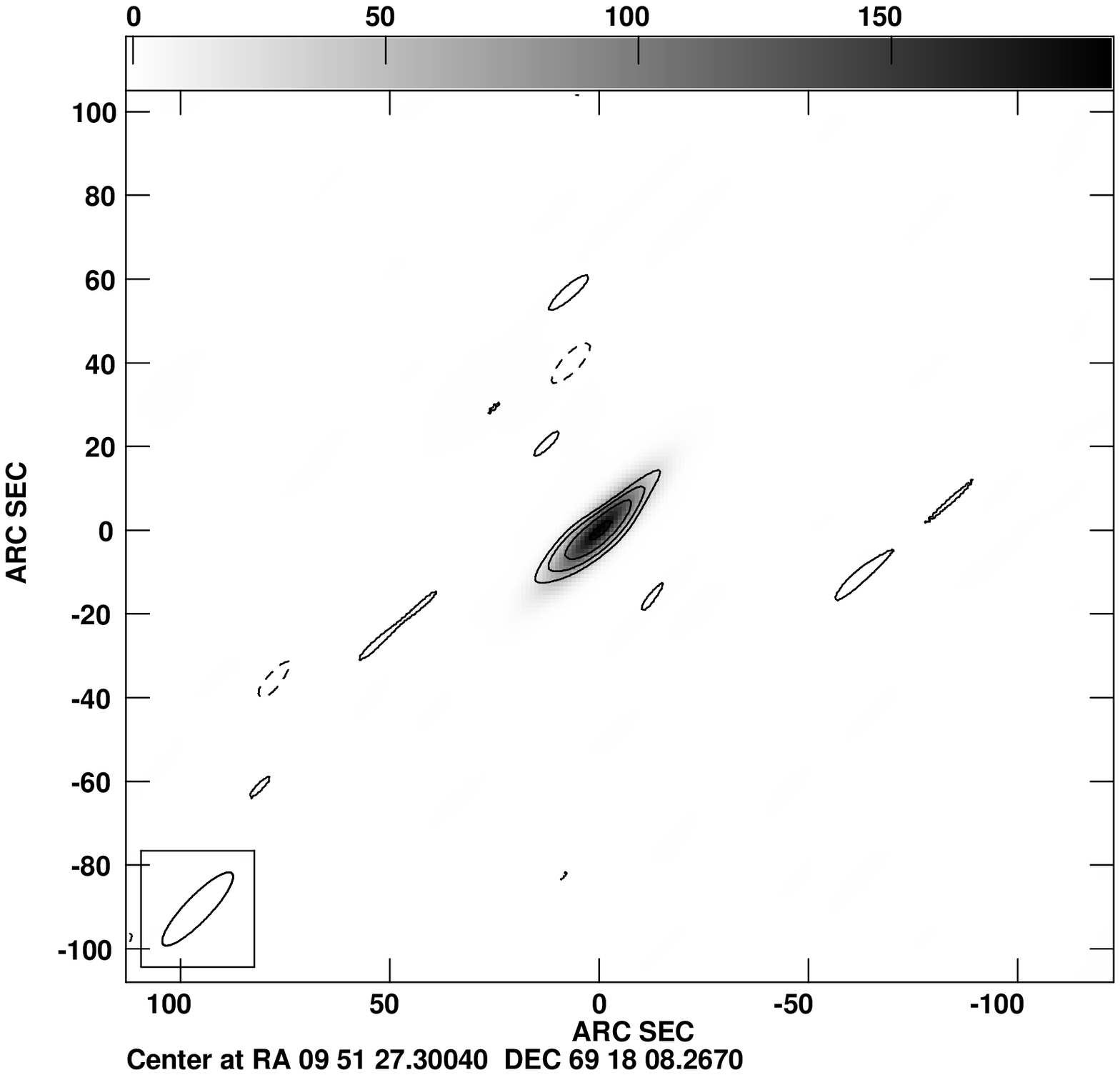}
\end{figure}
\figcaption[fig1.ps]{The Stokes V (contours) and total intensity 
(grey scale) map for M81 on 2000 September 27 at 8.4 GHz. The rms noise of the Stokes V map is 0.11 mJy beam$^{-1}$. The contours are 0.3*(-1,1,1.414,2,2.828,4,5.657) mJy and the peak flux density is 0.89 mJy/beam. \label{fig1}}

\end{document}